\documentclass[12pt]{iopart}
\pdfoutput=1
\usepackage{graphicx}
\usepackage{dcolumn}
\usepackage{bm}
\usepackage{float}
\usepackage{SIunits}

\usepackage{ifpdf}

\begin{document}

\title{ \bf Storage and Manipulation of Light Using a Raman Gradient Echo Process}

\author{M. Hosseini, B. M. Sparkes, G. T. Campbell, P. K. Lam, B. C. Buchler}

\address{Centre for Quantum Computation and Communication Technology, Department of Quantum Science, The Australian National University, Canberra, Australia}
\begin{abstract}
The Gradient Echo Memory (GEM) scheme has potential to be a suitable protocol for storage and retrieval of optical quantum information. In this paper, we review the properties of the $\Lambda$-GEM method that stores information in the ground states of three-level atomic ensembles via Raman coupling.  The scheme is versatile in that it can store and re-sequence multiple pulses of light. To date, this scheme has been implemented using warm rubidium gas cells. There are different phenomena that can influence the performance of these atomic systems. We investigate the impact of atomic motion and four-wave mixing and present experiments that show how parasitic four-wave mixing can be mitigated. We also use the memory to demonstrate preservation of pulse shape and the backward retrieval of pulses.

\end{abstract}

\maketitle

\section{Introduction}

Coherent storage and manipulation of light has attracted a lot of attention in the last decade due to its application in optical quantum communication  \cite{Gisin:RevModPhys:QC:2002} and quantum computation \cite{Monroe:PRL:Qgate:1995}.
There exist various light storage protocols based on coherent interaction of light with atomic ensembles. These include electromagnetically induced transparency (EIT) \cite{EIT-Lukin-PRL}, the atomic frequency comb (AFC)  \cite{AFC1,Clausen:2011p12253,Tittle:AFC:Nature2011}, four-wave mixing (FWM) \cite{Boyer:PRL:2007}, Raman absorption \cite{Reim:PRL:2011}, Faraday interaction \cite{Polzik:NatPh:2010}, controlled reversible inhomogenous broadening (CRIB) \cite{Nilsson:2005p4548,Kraus:2006bj}, and modified photon echo \cite{Damon:3PE:NJP:2011}.

The principle behind CRIB is to construct a reversible absorption process.  After absorption in an inhomogenously broadened atomic medium, some mechanism can be used to invert the detunings of the individual absorbers that are spatially distributed in the medium.  This reversal in detuning gives rise to a photon echo and thus the retrieval of stored optical information. CRIB was proposed in gas cells \cite{Moiseev:2001p4503} and solids \cite{Nilsson:2005p4548,SAMoiseev:2003p4515}, and first demonstrated in 2006 using a cryostatic ensemble of two-level rare-earth ions \cite{Alexander:2006p7654}.  It was then realized in 2008 that the application of a detuning gradient longitudinally along the length of the storage medium allowed for high recall efficiencies in the forward direction \cite{HetetPRL:100:2008} by preventing reabsorption of the light. This overcame a limit of 54\% recall efficiency that had previously been determined~\cite{Sangouard:2007cw}.  The use of a longitudinal gradient to control the broadening of the atomic ensemble is referred to as longitudinal-CRIB or gradient echo memory (GEM).  In 2010, ensembles of cryostatic rare-earth ions were used in a GEM system to demonstrate the first unconditional quantum memory with 69\% recall efficiency \cite{Hedges:2010p11910}.

The focus of this paper is $\Lambda$-GEM, which works using a three-level '$\Lambda$' atomic transition. It was initially implemented in a warm Rb vapor cell in 2008 \cite{Hetet:2008p5840}. This scheme has been used as a random access memory for light pulses, \cite{Hosseini:2009p8466}, has a demonstrated efficiency as high as  87\% \cite{Hosseini:NComm:2011}, and provides noiseless storage of quantum states \cite{Hosseini:NatPhys:2011}.

In this paper we review the basics of the $\Lambda$-GEM scheme and its description as a polariton in the spatial Fourier domain.  We discuss the particular issues that relate to the gaseous rubidium vapor that is the basis of our experiments, in particular the effects of diffusion. We present experiments showing the forward and backward retrieval of stored pulses as well as pulse shape preservation. We then model the impact of four-wave mixing in our system, which is a potential source of noise for stored quantum states. Our work shows, both through models and experiments, that we can mitigate the effects of four-wave mixing on the $\Lambda$-GEM scheme.

\section{Light storage using GEM}
  
\begin{figure}[!h]
  \centering
  \includegraphics[width=\columnwidth]{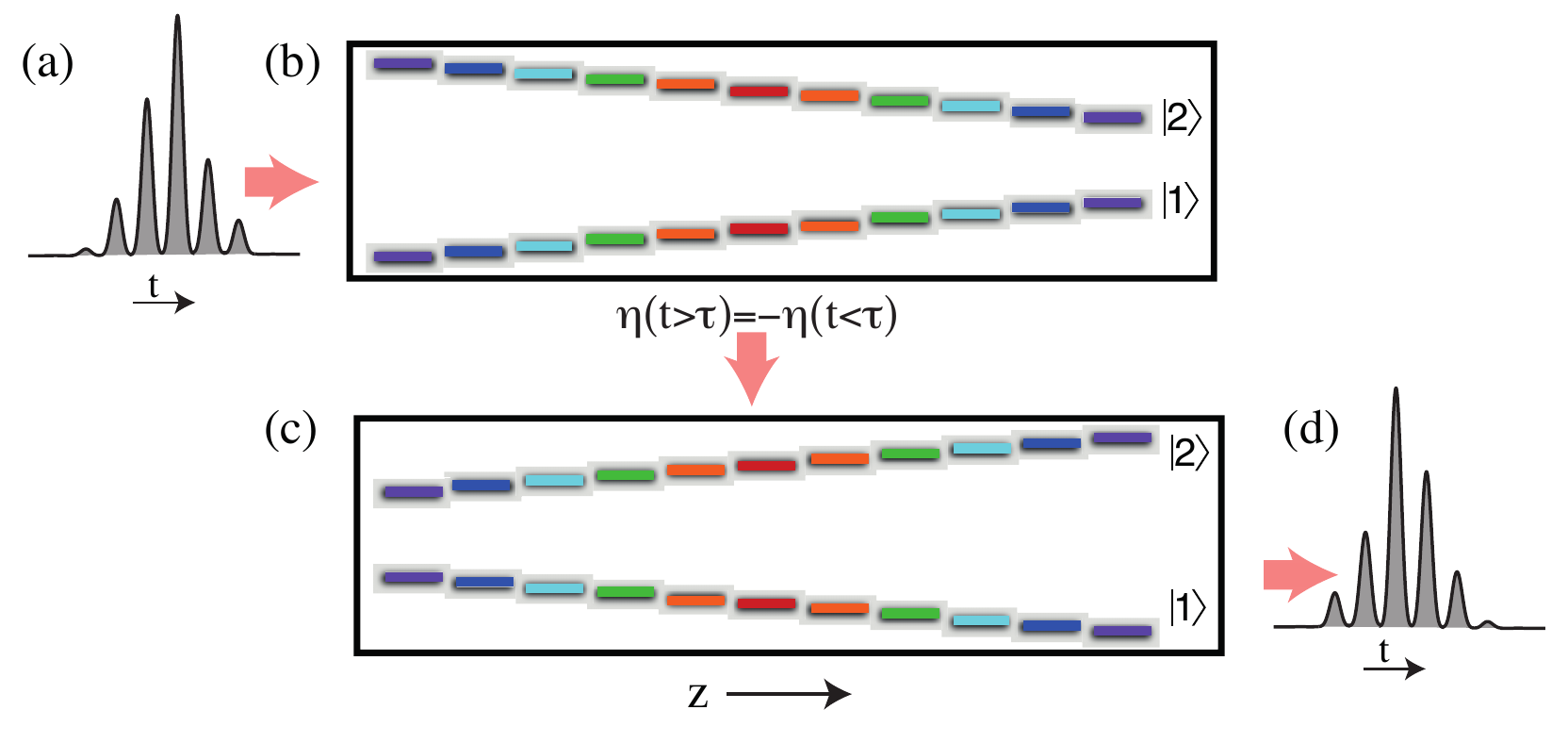}
 \caption{Schematic representation of a two-level GEM. An intensity modulated light pulse, (a), is incident on an atomic medium, (b), with an atomic frequency spectrum ($\eta z$). The bandwidth of the atomic frequency gradient covers the input pulse spectrum so that each frequency component is stored at different position along the memory. To release the pulse, $\eta$ is switched to $-\eta$, (c), after a time $\tau$. The stored light emerges as an echo, (d), in the forward direction at $t=2\tau$.}
  \label{schem}
 \end{figure}

Consider an ensemble of two-level atoms, as shown in Fig.~\ref{schem}. Due to the atomic frequency gradient, each frequency of the input optical pulse (probe) will absorbed by resonant atoms at different points along the length of the ensemble.  The atomic polarization along the  $z$ direction is thus proportional to the Fourier spectrum of the input probe light. To release the stored light, the gradient $\eta$ is simply inverted at time $\tau$ and the optical field is regenerated as a photon echo at time $2\tau$. The echo pulse emerges in the forward direction but shape-reversed. 
 
 The GEM scheme relies on inhomogeneous broadening being introduced as a linear atomic frequency gradient, $\delta(z,t)=\eta(t) z$, along the length of the storage medium, $0<z<L$, where $\eta(t)$ is the slope of the gradient that can change in time. In theory, GEM can reach 100\% efficiency in the limit of high optical depth (OD) \cite{HetetPRL:100:2008,Longdell08pra}. 
 
 Two-level GEM has been demonstrated using a cryogenically cooled solid state system. A storage efficiency of 69\% for coherent states has been demonstrated \cite{Hedges:2010p11910}. Furthermore, it was shown that the noise introduced by the memory is negligible for very weak coherent states. Using optical transitions, however, means that the storage time is limited by the excited state linewidth. Below, we introduce the $\Lambda$-GEM scheme that, in principle, offers high efficiency, long storage time, and extra control over manipulation of the optical information.
 
\begin{figure}[!h]
  \centering
  \includegraphics[width=0.6\columnwidth]{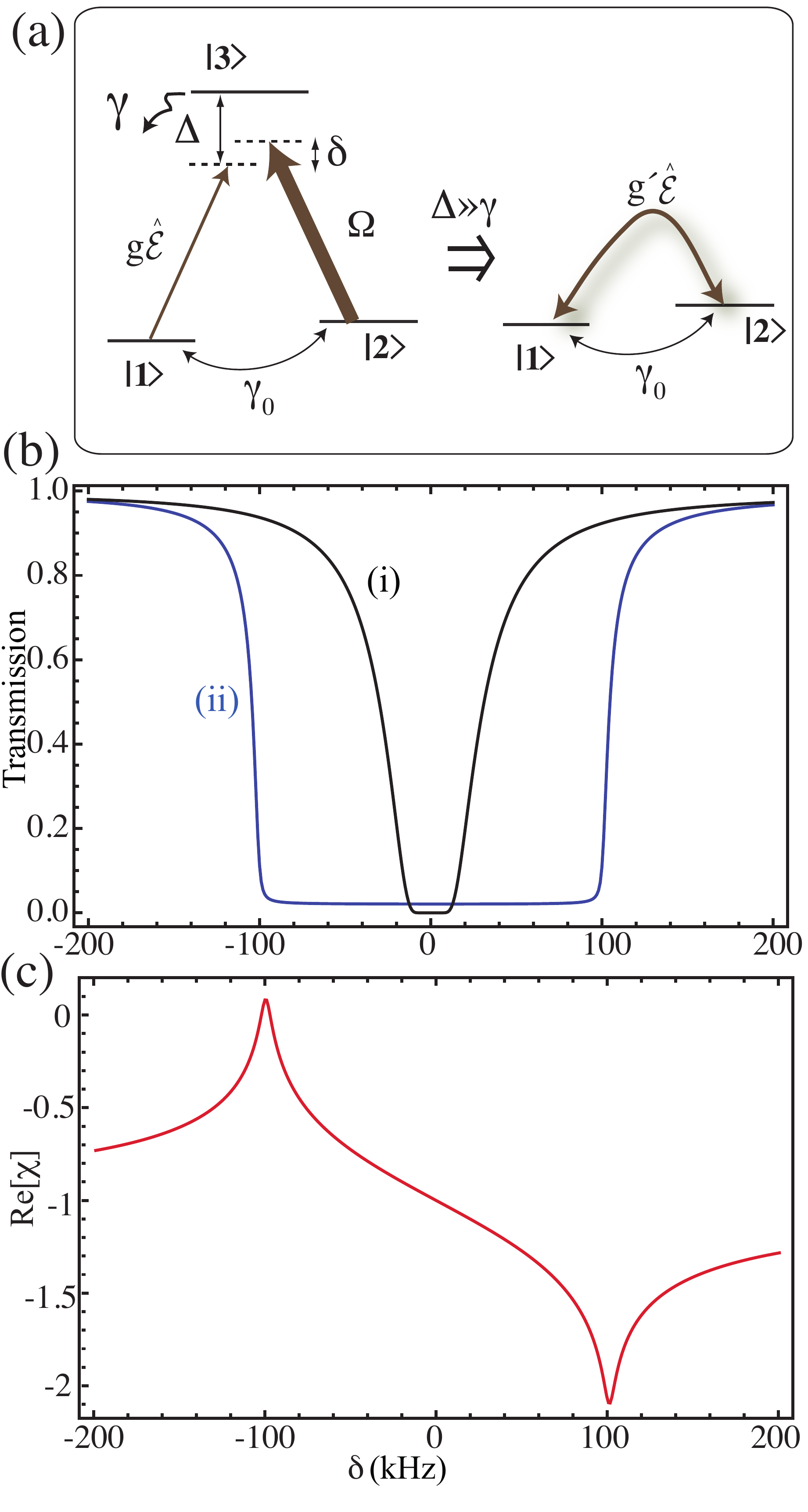}
 \caption[$\Lambda$ atom interacting with two laser fields]{(a) Atomic level structure of a $\Lambda$-type transition with two laser fields (coupling and probe fields) with Rabi frequency of $\Omega$ and $g\hat{\mathcal{E}}$, respectively, where $\Omega\gg g\hat{\mathcal{E}}$. The atomic spin decoherence rate is $\gamma_0$, one-photon and two-photon detunings are $\Delta$ and $\delta$, respectively. The excited state linewidth is $\gamma$. In the limit of large detuning the excited state can be adiabatically eliminated and the $\Lambda$-atom (atom on the left) is equivalent to a quasi two-level atom, as shown on the right. (b) Normalized transmission of the probe signal as a function of two-photon detuning for unbroadened (i) and broadened (ii) Raman lines. (c) The real part of susceptibility corresponding to the broadened Raman line. Parameters used to calculate susceptibility are: $\Omega=20$ MHz, $\gamma_0=5$ kHz, $\Delta=2$ GHz, and $\eta L=200$ kHz for the broadened line.}
  \label{REIT}
 \end{figure}

Let us now consider the three-level structure depicted in Fig.~\ref{REIT} (a) where a strong coupling field and a weak probe field, with Rabi frequencies of $\Omega$ and $g\hat{\mathcal{E}}$, respectively, interact with a three level atom with one photon detuning $\Delta$ and two-photon detuning $\delta$. In the weak probe regime, we can assume that once all the population is pumped to the state $|1\rangle$ ($\sigma_{11}\simeq1$) it stays there during the interaction time. The Heisenberg/Maxwell equations can be written as

\begin{eqnarray}
\dot{\hat{\sigma}}_{13}  &=& -(\gamma+\gamma_0/2+i\Delta)\hat{\sigma}_{13} + i g
\hat{\mathcal{E}}+ i\Omega \hat{\sigma}_{12} + \hat{F}_{13} \nonumber  \\
\dot{\hat{\sigma}}_{12} &=&  -(\gamma_{0}+i\delta(z,t)) \hat{\sigma}_{12}+ i \Omega^{\ast}
\hat{\sigma}_{13} + \hat{F}_{12}  \label{eq:3lMB}\\
(\frac{\partial}{c\partial t}+\frac{\partial}{\partial z})\hat{\mathcal{E}}&=&i \mathcal{N}\hat{\sigma}_{13} \nonumber
\end{eqnarray}

where $\mathcal{N}=gN/c$ is the effective linear atomic density. The Langevin operators $\hat{F}_{13}$ and $\hat{F}_{12}$ account for noise arising from spontaneous emission $\gamma$ and ground state decoherence $\gamma_0$ respectively. It has been demonstrated that in the linear weak-probe regime, this system does not show any excess noise beyond that required the preserve commutation relations \cite{Hetet:EIT:PRA2008}.  In other words, the memory can only introduce vacuum noise associated with loss - just like a beamsplitter. This means we need not treat the noise terms with a full quantum model. Rather, we can run a semi-classical model of  the system using the equations without the Langevin noise, and then add the vacuum associated with the modeled efficiency. In future equations we will omit the Langevin terms and the hats from the operators and consider semi-classical c-number equations that we can solve numerically.

Using the steady state solution of Eqs.~\ref{eq:3lMB}, the susceptibility of the medium can be written \cite{Fleischhauer:RevEIT:2005} as 
  \begin{eqnarray}
\chi=i \frac{(8 \delta^2 \gamma +2 \gamma_0  (\Omega^2 + \gamma_0 \gamma)}{
 |\Omega^2 + (\gamma+ 2i \Delta) (\gamma_0 + 2i \delta )|^2}+ \frac{(4 \delta(\Omega^2 -4\delta\Delta)-4\Delta\gamma_0^2}{
 |\Omega^2 + (\gamma+ 2i \Delta) (\gamma_0 + 2i \delta )|^2.}
\label{eq:susc}
\end{eqnarray}
The normalized transmission of the probe field as well as the imaginary part of the susceptibility is depicted in Fig.~\ref{REIT} as a function of the two-photon detuning, $\delta$. At large ODs, the absorption of the unbroadened line (trace (i)) is significant and applying gradient-broadening of 200 kHz to the narrow absorption feature does not drastically alter the maximum absorption (trace (ii)). The real part of the susceptibility, shown in Fig.~\ref{REIT}(c), shows that there is substantial dispersion of probe field across the absorption feature.
 
 We now simplify the equations further by performing an adiabatic elimination of the excited state \cite{Hetet} and assuming $\partial\hat{\sigma}_{13}/\partial t\ll \gamma $, or equivalently $1/T \ll \gamma$, where $T$ is the fastest time-scale of the evolutions. We also assume a large detuning compared to the spontaneous emission rate ($\Delta\gg\gamma$). Assuming the coupling beam to be real and combining Eqs.~\ref{eq:3lMB} we reach
\begin{eqnarray}
\dot{\sigma}_{12}=(-\gamma_0+i\delta(z,t)-i\frac{\Omega^2}{\Delta}) \sigma_{12}- i \frac{g\Omega}{\Delta}\mathcal{E}    \\
(\frac{\partial}{c\partial t}+\frac{\partial}{\partial z})\mathcal{E}= \frac{ig \mathcal{N}}{\Delta}\mathcal{E}+ i \frac{\mathcal{N}\Omega}{\Delta}\sigma_{12}.
\label{eq:8}
\end{eqnarray}
The term $\Omega^2/\Delta$ is the ac-Stark frequency shift caused by the coupling field, which can be cancelled by adjusting the coupling field frequency. Performing the transformation $\mathcal{E}\to\mathcal{E}e^{ig \mathcal{N}c/\Delta t}$ and  $\Omega\to\Omega e^{-ig \mathcal{N}c/\Delta t}$ we can remove the first term on the right hand side of the Maxwell equation. We also perform the transformation $z'=z+ct$, i.e.  moving to a frame with speed of light, to reach
\begin{eqnarray}
\dot{\sigma}_{12} &=&  -(\gamma_0+i\delta(z'-ct,t))\sigma_{12}- i \frac{g\Omega}{\Delta}\mathcal{E}\label{eq:three1}    \\
\frac{\partial}{\partial z'}\mathcal{E}&=& i \frac{\mathcal{N}\Omega}{\Delta}\sigma_{12}
\label{eq:three2}
\end{eqnarray}
The equations for two-level GEM are \cite{HetetPRL}
 \begin{eqnarray}\label{ANALGEM}
\dot{\sigma}_{12} &=&  -(\gamma_{12}+i\delta(z,t))\sigma_{12}- ig\mathcal{E} \label{eq:2latom1}   \\
\frac{\partial}{\partial z}\mathcal{E}&=& i\mathcal{N}\sigma_{12}
\label{eq:2latom2}
\end{eqnarray}
which are formally equivalent to our Raman scheme if we set 
 $\mathcal{N}\to\mathcal{N}\Omega/\Delta$,  $g\to g\Omega/\Delta$ and $\gamma\to\gamma_0$. Therefore, one can see that the GEM storage mechanism described above for two-level atoms is also applicable to the three-level systems in certain regimes. The two levels are now the long-lived ground states of the three-level system.
 
  \section{Polaritonic description of GEM}
  
 It is possible to associate a quasi-particle (polariton) with the light field and spin wave inside the GEM system, similar to the EIT dark-state polariton \cite{EIT-Lukin-PRL}. The evolution of the polaritonic mode inside the GEM system provides a clear intuitive picture of the system. Neglecting the decoherence and taking the spatial Fourier transform of Eqs.~\ref{eq:2latom1} and \ref{eq:2latom2} we arrive at
\begin{eqnarray}
\frac{\partial}{\partial t}\sigma_{12}(t,k)&=&\eta\frac{\partial}{\partial k}\sigma_{12}(t,k)+ig\mathcal{E}(t,k)\\
k \mathcal{E}(t,k)&=&\mathcal{N}\sigma_{12}(t,k)\label{eq:kEeqNa}
\end{eqnarray}
and therefore
\begin{eqnarray}
(\frac{\partial}{\partial t} -\eta(t)\frac{\partial}{\partial k}-i\frac{g\mathcal{N}}{k})\sigma_{12}(t,k)=0
\end{eqnarray}
Here $k$ is the spatial frequency component of the joint system. A similar equation can be written for the electric field, $\mathcal{E}$. Thus, a single mode polariton-like operator in time and $k$ space for a two-level atom is obtained \cite{PRL-3GEM}, $\psi(k,t)=k\mathcal{E}(k,t) + \mathcal{N} \sigma_{12}(k,t) $, which has the following equation of motion,
 \begin{eqnarray}\label{eq-motion}
(\frac{\partial}{\partial t} - \eta(t)\frac{\partial}{\partial k} - i\frac{g \mathcal{N}}{k})\psi(k,t) &=&0. 
\label{eq:motion}
\end{eqnarray}

Another solution for the above equation of motion as $\psi(k,t)=k\mathcal{E}(k,t) - \mathcal{N}\sigma_{12}(k,t) $. Using Eq.\ref{eq:kEeqNa}, it is evident that the anti-symmetric mode will never be excited. The normal mode equation of motion (Eq.~\ref{eq:motion}) indicates that ${\psi}(k,t)$ propagates along the $k$-axis with a speed defined by the slope of the gradient. As the polariton reaches higher $k$ values the electric field amplitude decreases. As each frequency component of light propagates through the medium and gets absorbed by resonant atoms, the dispersion due to the neighboring atoms, having only slightly different resonant frequencies, will affect the propagation and absorption of the light.  This is, in fact, the origin of the photonic component of the polariton and also the phase shift of the echo investigated in Ref.~\cite{Moiseev:PRA:2008}. The intensity of the atomic excitation, created after the input light enters the medium, remains unchanged during the storage time. Flipping the gradient (i.e. $\eta \rightarrow -\eta$) time reverses the absorption process so that when the polariton reaches $k=0$ (the phase matching condition) a photon echo emerges from the ensemble in the forward direction. We note here that at any time, the system is described by a normal mode with different amplitudes for optical and atomic fields, therefore it makes more sense to picture the GEM storage as continuum of evolving polaritons rather than just a single polariton.  

In a three-level atomic ensemble, in the far-detuned and adiabatic regime, using Eqs.~(\ref{eq:three1}) and (\ref{eq:three2}) a similar polaritonic mode can be described as $\psi(k,t)=k\mathcal{E}(k,t) + \mathcal{N'}\sigma_{12}(k,t) $, where $\mathcal{N'}=\mathcal{N} \Omega/ \Delta$. The second term in this equation necessarily goes to zero if the coupling field is switched off during storage. Therefore, as it is still the case that $k\mathcal{E} = \mathcal{N'}\sigma_{12}$, the slowly propagating electric field vanishes and the polariton becomes purely atomic.

\begin{figure}[!hb]
\centerline{\includegraphics[width=0.4\columnwidth]{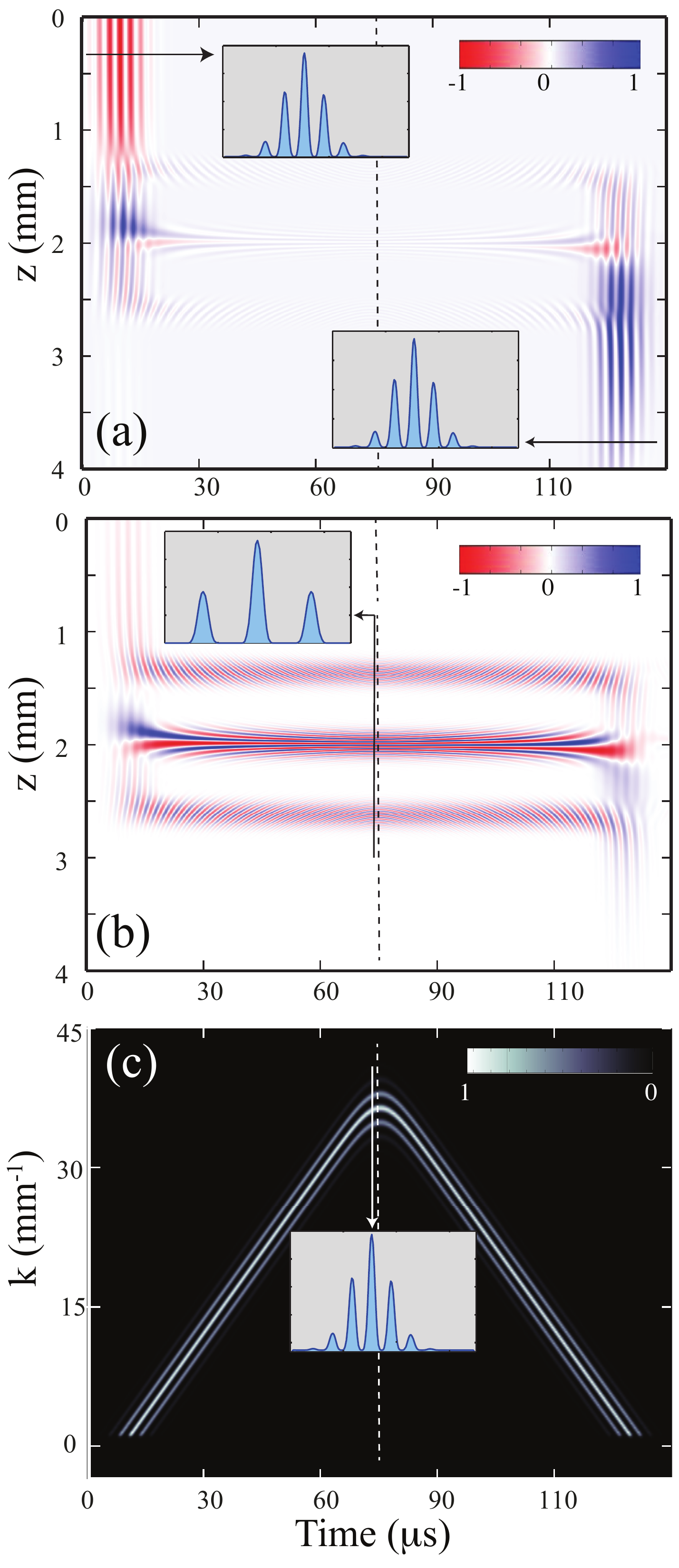}}
 \caption[GEM normal mode]{ Numerical simulation of the storage of an intensity modulated pulse. (a)~The real part of the electric field in $z-t$ plane. Inset shows the temporal shape of the pulse at the input and at the output. The electric field decreases in strength until the atomic frequency gradient is switched at $t=75$ $\micro s$. It then increases and eventually an echo is emitted. (b)~The real part of the atomic polarization in the $z-t$ plane. A cross-section of the atomic polarization along the z-axis reveals the Fourier spectrum of the input probe light. The inset shows the absolute value of the polarization at the indicated position. (c) Intensity of the polaritonic excitations in the $k-t$ plane. $k=0$ indicates that the rephasing of the atomic spin is completed to produce a coherent emission. A cross section of the polariton along $k$ reveals the temporal shape (amplitude) of the stored pulse (as shown in the inset). Parameters used for these simulations are: $g\mathcal{N}/\eta=1.74$, $\eta L=16\gamma$, and $\gamma_0=0$.}
 \label{pol}
 \end{figure}

The real parts of the electric field and atomic polarization, for a modulated input pulse, are plotted in Fig.~\ref{pol} (a) and (b), respectively, obtained using an XMDS \cite{xmds} numerical simulation of Eqs.~(\ref{eq:2latom1}) and (\ref{eq:2latom2}). It can be seen that the light is nearly stopped at the centre of the ensemble after the pulse enters the medium and its intensity gradually decreases. The atomic spin wave is quickly generated after the pulse enters the memory. We also see that the spatial structure of the atomic and electric fields becomes increasingly fine as time progresses before gradient switching.

The spatial cross section of the atomic polarization at any time during the storage is the Fourier spectrum of the input pulse. This explicitly demonstrates the frequency-encoding nature of GEM. Conversely, any cross-section of the polariton $\psi(k,t)$ along the $k$ axis shows the temporal profile of the pulse \cite{PRL-3GEM} as a second Fourier transform into $k$-space returns the original pulse shape. This is shown in Fig.~\ref{pol} (c) where the polaritonic normal mode of the modulated pulse is plotted.

It is quite interesting to compare the polariton pictures in EIT and GEM systems~\cite{PRL-3GEM}. The EIT polariton propagates through the medium (in $z-t$ plane) while the control field is on and becomes stationary when control field is switched off. Thus, the temporal modes compressed within the atomic sample and, by integrating the polariton amplitude along the propagation axis, one can construct the temporal shape of the light field. The GEM polariton, on the other hand evolves in the spatial frequency domain and its evolution can be controlled via the external field, as described above.   

 The group velocity of the light field propagating inside the GEM medium is given  by $v_g=g\mathcal{N}/k^2$. The further away the polariton is from $k=0$, the less intense the electric field and the smaller the group velocity becomes. In the $\Lambda$-GEM scheme, the group velocity and light amplitude can be controlled by both the coupling field and the gradient field. This extra control over the information provides considerable flexibility for data manipulation.

 \subsection{Steering of the GEM polariton}
  
 The dynamic characteristics of the GEM polariton allow one to precisely control and manipulate the state of the system  in time. This control is even more versatile in $\Lambda$-GEM due to the contribution of the coupling field and detuning gradient in determining the state of the system. 
 
 \subsubsection{Atomic detuning.}
 
The detuning gradient causes the different frequency components of the light to be absorbed spatially along the atomic memory. In the polaritonic picture, the phase matching condition is equivalent to $k=0$. After the light is stored, $\eta(z)$ can be switched to any arbitrary shape in time. The photon echo, however, is only emitted when the atoms are rephased and coupling field is on. The shape of the frequency detuning is chosen to be monotonic in order to avoid any re-absorption. One can use other forms of frequency detuning to avoid reabsorption, but a linear detuning is practically easier to implement and control. 

The shape of the detuning can be engineered in such a way as to optimize storage of a light pulse with a particular frequency spectrum. The size of the frequency gradient determines the bandwidth of the memory and, by applying a larger gradient field, the storage bandwidth increases. The number of well separated pulses, each with a bandwidth smaller than the memory bandwidth, that can be simultaneously stored is limited by the memory lifetime. For a simple linear gradient, the efficiency of GEM is inversely proportional to the size of the gradient. There will, therefore, be a trade-off between the bandwidth and storage efficiency. For efficient storage of broadband information, large ODs are imperative.

\begin{figure}[ht]
\centerline{\includegraphics[width=\columnwidth]{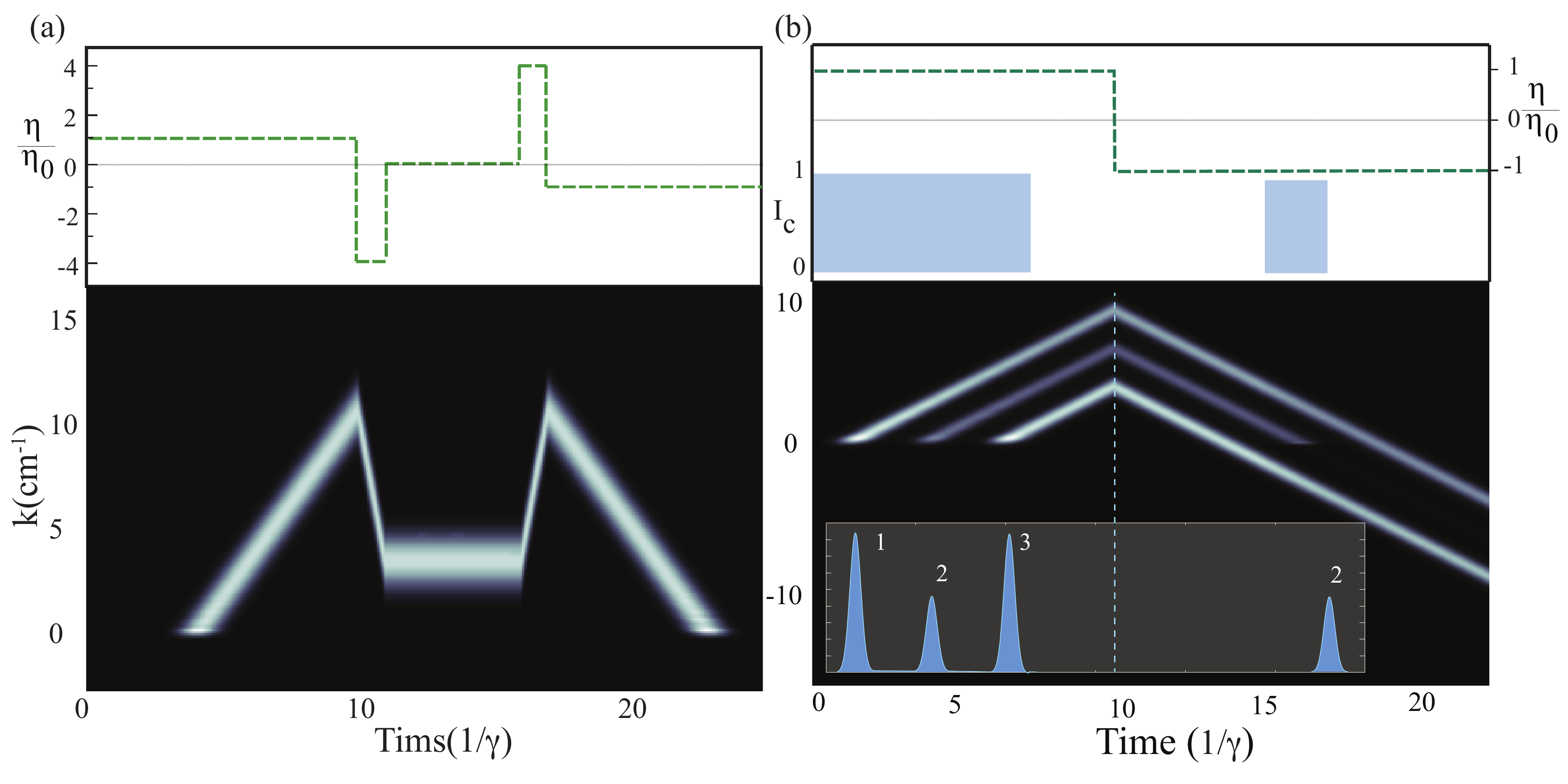}}
 \caption[Switching of GEM polariton]{Numerical simulation of GEM showing absolute value of the polaritonic excitation in $k-t$ plane. (a) Multiple switches of the atomic gradient field. The switching protocol of the gradient detuning is depicted on top where $\eta$ is switched to $\eta=-4\eta_0$, then $\eta=0$, $\eta=4\eta$ and finally to $\eta=-\eta_0$. (b) Arbitrary probe retrieval. The coupling field is turned on during the writing stage and only for a short period during the reading stage when the condition $k=0$ for the middle pulse is satisfied. This guarantees that only the second pulse is recalled. The top part of the figure shows the switching protocols for coupling field intensity , $I_c$, and detuning gradient, $\eta$. The inset shows the temporal profiles of the input and echo pulses.}
 \label{swzc}
 \end{figure}

The frequency gradient can be switched to higher or lower values or even to zero during the storage time without affecting the storage mechanism. In Fig.~\ref{swzc} we show numerical simulations for the evolution of the normal mode in the $k-t$ plane where $\eta$ is switched to different values. In Fig.~\ref{swzc} (a), a recall gradient four times steeper than the input gradient introduced and therefore the atomic field evolves faster towards the origin. Shortly after, before the polariton reaches $k=0$, $\eta$ is switched to zero to stop the evolution and freeze the polariton at a constant $k$ value. The system can then be described as a single polariton. Eventually, with $\eta = -\eta_0$, light is emitted when atomic excitations are phase-matched at $k=0$.

\subsubsection{Tunable atom-light coupling.}

The role of the classical coupling field in $\Lambda$-GEM is to couple the weak probe light via a Raman transition to the ground states. The coupling field is also required at the reading stage when the photon echo is emitted to transfer the information from the atomic spin back to the light field. In an ideal system, the dynamics of the coupling field during the storage time would not affect the storage process, but practically it is beneficial to turn off the coupling field during the storage time to reduce scattering and therefore loss. The scattering process is further discussed in the experiment section.  

 The coupling field intensity determines the effective atom-light coupling and therefore the OD. Decreasing the coupling field intensity reduces the storage efficiency at the writing stage. At the reading stage this can result in partial retrieval; the intensity of the coupling field when the normal mode reaches $k=0$ determines how much stored excitation is converted to the electric field and how much is left inside the memory.
 
 Fig.~\ref{swzc}(b) shows a simulation of the normal mode evolution in the $k-t$ plane in the $\Lambda$-GEM system. The coupling field intensity is switched to zero shortly after the three light pulses enter the medium and their information is transferred into the atomic excitations. After the gradient is flipped and the excitations reach $k=0$, the coupling field is turned back on only for the time window  $t=16$ to $18 \gamma^{-1}$ in order to couple out the second pulse. In this case, the information imprinted into the atomic ensemble from the two other pulses will remain inside the memory.
 
\section{Time sequencing and manipulation of pulses}

Using the polariton description introduced above, we show here how optical information can be coherently manipulated in an arbitrary manner. We discuss the possibility of coherent pulse sequencing, with first-in-last-out (FILO) , first-in-first-out (FIFO), arbitrary recall and backward retrieval of information using $\Lambda$-GEM \cite{Hosseini:2009p8466}.
  
  \subsection{First-in-last-out storage}
  
One of the intrinsic properties of GEM is that the photon echo is a time-reversed copy of the input pulse. When the optical field frequencies are stored spatially along the z-axis, flipping the gradient triggers the time-reversal process so that the pulse sequence is reversed.

Fig.~\ref{fiao} (a) shows numerical simulation of the polariton for first-in-last-out (FILO) storage of four input pulses, depicted in $k-t$ plane. The first pulse to enter the system returns to $k=0$ last, and is thus re-emitted last.  
  \subsection{First-in-first-out storage}

\begin{figure}[!h]
\centerline{\includegraphics[width=0.6\columnwidth]{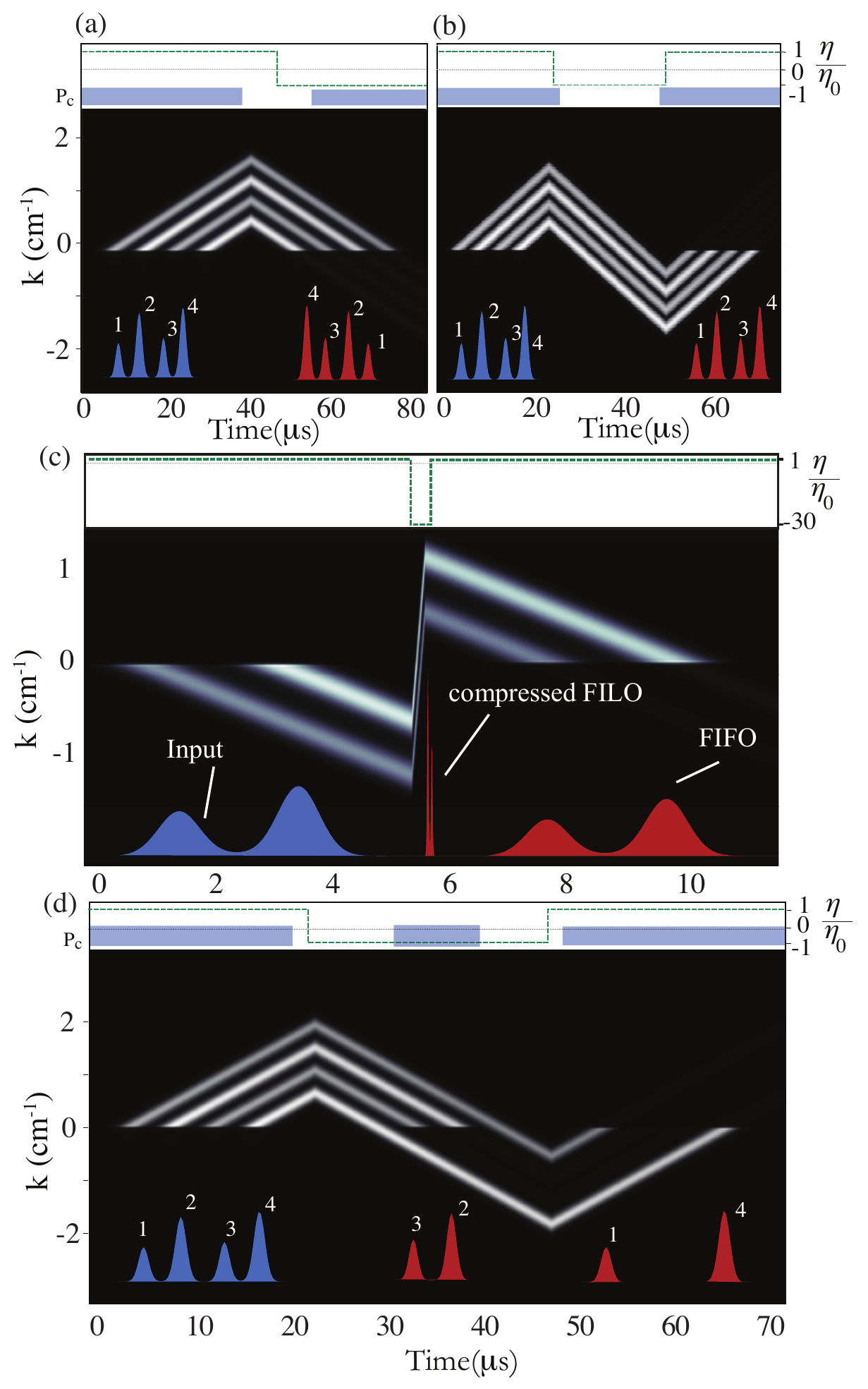}}
 \caption[First in first out storage]{ Numerical simulation of absolute value of the polaritonic excitations showing, (a) first-in-last-out and (b) first-in-first-out in a three-level system of four input pulses in $k-t$ plane. (c) FIFO memory in a two-level system. The original gradient is switched to a much steeper negative gradient to reduce the effective OD of the memory. The two pulses are retrieved in FIFO manner after switching the gradient once more to its original slope. In this case, around 7\% of the light is retrieved as FILO at the first readout stage. (d) Storage of a train of four pulses in 1,2,3,4 order and recalled in  3,2,1,4 order. The switching scheme of the coupling field intensity and gradient detuning is plotted on top of each part. Insets show the temporal shape of input (blue) and echo (red) pulses. }
  \label{fiao}
 \end{figure}

 Shape preserving storage in GEM is possible if the absorption is time-reversed twice. The first pulse in would then be the first pulse out. Using a three-level system, one is free to turn off the coupling beam. In this case, although the atomic polarisation is completely rephased when the normal mode reaches $k=0$,  no light can be emitted because the atom-light coupling is turned to zero. This is seen in Fig.~\ref{fiao}(b), which shows $|\psi(t,k)|^{2}$ for the switching scheme shown on the top part of the figure. With the coupling beam off, the normal mode passes straight through $k=0$ to negative $k$ values. We can then switch the frequency gradient once again to time-reverse the atomic spins one more time. Now with the coupling field back on, the normal mode is converted into a photon echo at $k=0$ without pulse shape reversal. In this way, we can construct a first-in-first-out (FIFO) memory.
 
 FIFO storage can also be achieved in a single memory operation based on two-level GEM, albeit with some loss. This can be done by switching the gradient to a very steep value and opposite sign after storage as shown in Fig.~\ref{fiao} (c). This will effectively reduce the OD of the system at the first reading stage and therefore most of the excitation will remain inside the memory \cite{Longdell:2008p8530}. The left-over excitations, that contain most of the information, can be recalled later by applying a gradient similar to the one at the writing stage.
  
  \subsubsection{Multiple retrievals}

In the two-level GEM scheme, for low OD regimes, the writing stage becomes inefficient as light will leak through the atomic ensemble without being absorbed. The fraction of the light that is absorbed is $1-\exp(-2 {\beta} {\pi})$, where $\beta=g\mathcal{N}/\eta$. Similarly, at the read stage, low OD means that not all of the atomic polarization will be converted back to optical field.  A fraction $\exp(-2 {\beta} {\pi})$ of the atomic excitation remains trapped in the memory \cite{Longdell:2008p8530}. Provided that the coherence time of the atoms is long enough, part of the left-over excitation can be converted back to a light field by flipping the field gradient multiple times. In fact, multiple switching of the gradient allows atomic excitations to travel back and forth in $k$ space, and each time they pass through $k=0$, a fraction of excitation can contribute to photon echo emission until, eventually, all the left-over excitation becomes  depleted. Disregarding decoherence, the fraction of the input light released in the n$^{th}$ echo is $[1- \exp(-2 \beta\pi)]^2 \exp[-2 \beta\pi (n-1)]$. 

In the case of the $\Lambda$-GEM scheme, however, the presence of the coupling field provides another degree of freedom to control the amount of atomic excitation converted to the light after each gradient flip. By leaving the coupling field off when $k=0$, the emission is suppressed and the atoms hold on to the information. Partial retrieval can be achieved by lowering the coupling field power at the readout. This ability can be used to create a time-delayed beam splitter or polariton-polariton interferometer \cite{Campbell:arxiv:2011}.

  \subsection{Arbitrary retrieval.}

By controlling the frequency gradient and coupling field intensity in time, it is possible to construct a system that can recall the pulses in any arbitrary order. In the case of multi-pulse storage, one can decide to turn on the coupling field only during the chosen time windows, when atomic excitations are at $k=0$, to recall only desired bins of the stored information and leave the rest inside the memory. A decoherence-free model of the on-demand retrieval is shown in Fig.~\ref{fiao}(d), where 4 pulses are stored in the memory, and after the first field switch, pulses 3 and 2 are recalled by turning the coupling field on, only during the time window that these two pulses cross $k=0$. The other two pulses are recalled later after the second field switch. One can choose any other combination in the reading stage by controlling the detuning gradient and also the coupling field switching. 

\subsection{Backward propagating retrieval of echo}

 Photon echoes created using two-level GEM always co-propagate with the input pulse. The $\Lambda$-GEM is capable of recalling stored information in both backward and forward directions \cite{Carreno:GEM:2011}. By turning off the coupling field co-propagating with the signal field, the slow light vanishes, and a collective spin excitation is created. At the reading stage. when the coupling field is turned back on, the signal pulse is recreated, propagating along the quantization axis in the same direction as of the coupling field. If a coupling field counter-propagating with the initial signal field is applied at the read out stage a photon echo will be emitted in the backward direction. We demonstrate this below in the experimental section. It is also possible to create stationary light inside the memory by means of counter-propagating coupling fields, similar to EIT \cite{Lukin-EIT-Nature,Lin:PRL:2009}. This matter will be left for the future publications.

\section{Implementation in warm vapor}

We have experimentally implemented $\Lambda$-GEM using a $^{87}$Rb vapor cell at 70-80$^{\circ}C$. At such high temperatures the effects of Doppler broadening, diffusion, and collisions can impact the storage mechanism. Before presenting experimental data, we will discuss  some aspects of warm vapor that impact on the implementation of GEM in Rb vapor.   
\subsection{Angular dependency of absorption and transmission lines}
\begin{figure}[t!]
  \centerline{\includegraphics[width=0.8\columnwidth]{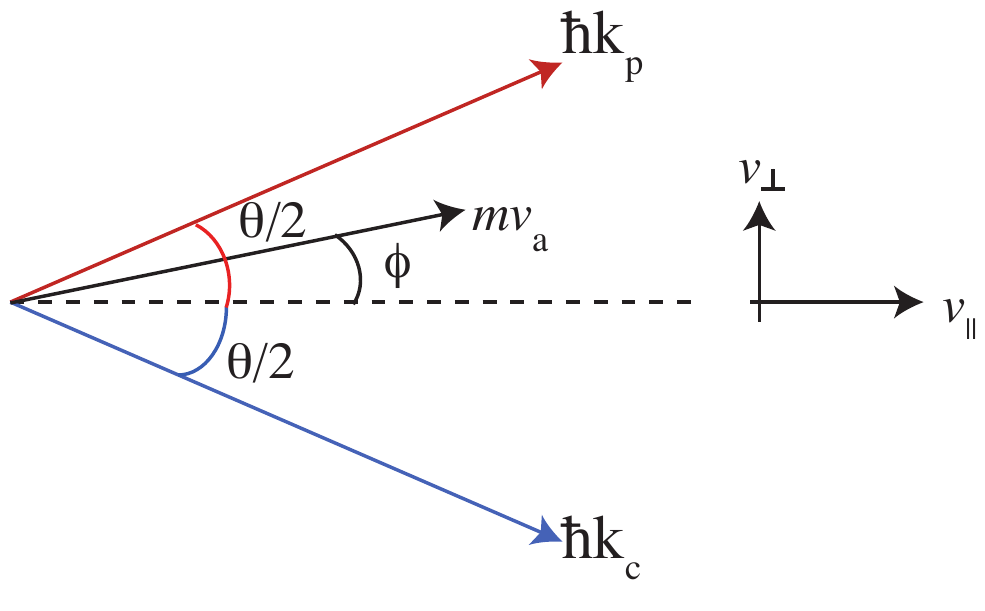}}
  \caption[Crossing angle between optical and atomic beams]{Schematic diagram showing crossing angle between optical and atomic beams. The reference coordinates are $v_{||}$ and $v_{\bot}$ representing quantization axis and axis perpendicular to it, respectively. $m$ and $v_a$ are the mass and velocity of the atom, $k_p$ is the wave-vector of the probe light and $k_c$ is the wave-vector of the coupling light.}
  \label{angle}
  \end{figure}

The mode-matching and crossing angle between the coupling and the probe beams are crucial for performance of both EIT- and Raman-based memories. 
To study this effect we consider the angular configuration of optical and atomic beams shown in Fig.~\ref{angle} where the angle between the control and probe is shown as $\theta$ and the atomic velocity makes an angle $\phi$ with the quantization axis. 

    \normalsize

 In a vapor cell, if the crossing angle between the beams is non-zero, the one-photon and two-photon detunings need to be modified to include extra frequency shifts \cite{Purves:PhD:2006} in order to describe the physics of the system more accurately. Change in the one-photon and two-photon detuning of the Raman line are given by 
 \begin{eqnarray}
\Delta\to\Delta -{\bf k_p.v}\nonumber \\ 
\delta\to\delta +({\bf k_p-k_c)}{\bf .v}
\end{eqnarray}
where 
\begin{eqnarray}
-{\bf k_p.v}&=& -k_p v_a \cos(\theta/2+\phi)= - k_p v_p\nonumber \\ 
({\bf k_p-k_c)}{\bf .v}&=& 2k_pv_{\bot}\sin(\theta/2)
\end{eqnarray}
and bold letters represent vectors. The velocity, $v_p$, is the component of atomic velocity along the probe beam and  $v_{\bot}$ is the component of atomic velocity perpendicular to the quantization axis. Taking these frequency shifts into account, the imaginary part of the susceptibility (see Eq.~\ref{eq:susc}) can be rewritten as
\begin{eqnarray}
Im[\chi]=\frac{(8 (\delta- k_p v_p)^2 \gamma +2 \gamma_0  (\Omega^2 + \gamma_0 \gamma)}{
 |\Omega^2 + (\gamma+ 2i (\Delta+k_pv_{\bot}\sin(\theta/2))) (\gamma_0 + 2i (\delta- k_p v_p) )|^2}
\label{eq:Imgsusc}
\end{eqnarray}

\begin{figure}[!h]
  \centerline{\includegraphics[width=0.6\columnwidth]{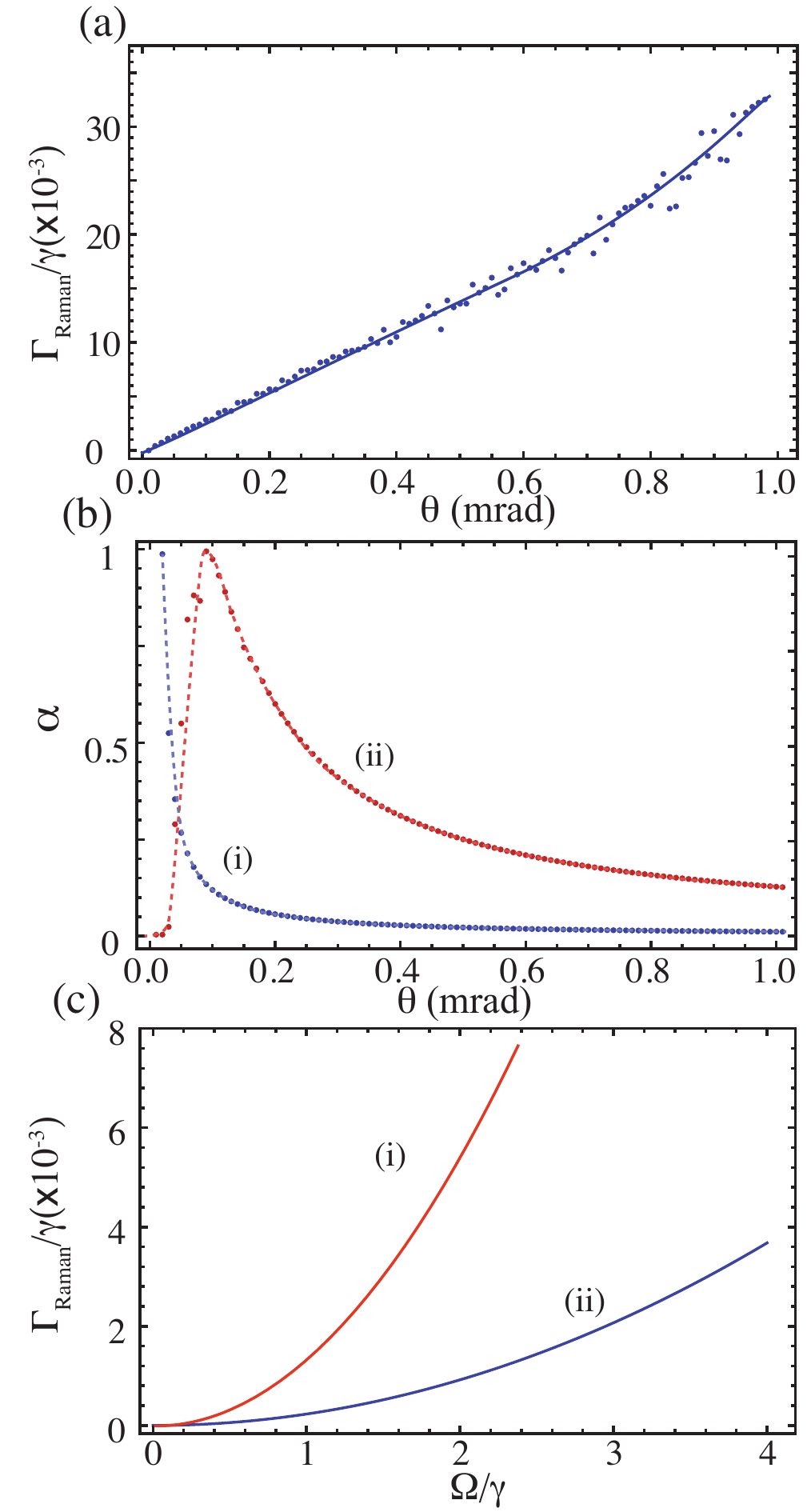}}
  \caption{ (a) FWHM of Lorentzian Raman absorption line calculated using Eq.~\ref{eq:int} where $\Omega=1.0\gamma$. The solid line is to guide the eye and fluctuation of the points is due to lack of precision in the integration. (b) Normalized absorption coefficient calculated for detuned light by numerical integration of Eq.~\ref{eq:int} for different crossing angle of coupling and probe beams. Red (i) and blue (ii) data show the absorption coefficient for $\Omega=1.0\gamma$ and $\Omega=2.0\gamma$ respectively.  The dashed lines are guides only. (c) FWHM of the Raman line plotted using numerical integration for $\theta=0$, (i) $\Delta=100\gamma$ and (ii) $\Delta=300\gamma$. Doppler effect has been included in the model and other parameters used to plot (a), (b) and (c) are: $\gamma_0=0$, $T=350$K. }
  \label{AngRaman}
  \end{figure}

At temperatures higher than room temperature, the atomic velocity in each direction is defined by the Maxwell-Boltzman distribution. We assume that the velocity is isotropic in the vapor cell, so the distribution of $v_z$ is the same as the distribution of $v_{\bot}$ and $v_{p}$. The absorption coefficient of the medium is then given by  
 \begin{eqnarray}
 \alpha\propto\int_{-\infty}^{\infty}\int_{-\infty}^{\infty}Im[\chi] (v_p,v_{\bot}) N(v_p)N(v_{\bot}) dv_pdv_{\bot},
 \label{eq:int}
\end{eqnarray}

where  $N(v)$ is the Maxwell-Boltzman distribution function. This integral can be numerically solved at $T>0$ to find the absorption coefficient and the Full-Width-at-Half-Maximum (FWHM) of  Lorentzian-shaped Raman lines. The Raman linewith shows a linear relationship with the crossing angle as shown in Fig.~\ref{AngRaman} (a). The absorption coefficient for two different values of $\Omega$, as a function of the crossing angle, is plotted in Fig.~\ref{AngRaman} (b). The crossing angle corresponding to the maximum absorption of the Raman line approaches zero at $\Delta/\Omega\gg1$. 

\subsection{Doppler broadening}

In order to describe the properties of warm vapor cells used in our experiment, we need to consider Doppler effects caused by the thermal motion of the atoms. The velocity spread of the contributing velocity classes can be roughly determined by taking into account the optical linewidth of the excited stated denoted by $\gamma$. The velocity spread leads to a Doppler broadening of the observed Raman line in the laboratory frame. This width at $T\simeq70^{\circ}$C is around $\Gamma_D=500$ MHz. 
 
  We follow a similar method as the one described in Ref. \cite{Javan:PRA:2002} and solve the integrals numerically to calculate the FWHM of the Raman line for different coupling field Rabi frequencies. The results of numerical calculation of the Raman linewidth for two different detunings as a function of $\Omega$ are depicted in Fig.~\ref{AngRaman} (c). We note here that the far-detuned two-photon absorption, i.e. $\Delta\gg\Gamma_D$, would be insensitive to the Doppler broadening of the excited state as long as the two fields are co-propagating. 


\subsection{Buffer gas in the vapor cell}

 In most of the light storage experiments that have been performed on warm Rb vapor cell, inert buffer gases such as Ne, Kr and He have been used to reduce the mean free path of the Rb atoms and therefore increase the time of flight of atoms inside the laser beams. The collision of Rb with buffer gas is elastic and therefore the phase information is better preserved, i.e. the ground state atomic spin remains largely unchanged. Collisions with buffer gas may, however, shuffle population in the excited state and can cause collisional broadening \cite{Erhard:DR:PRA:2001} that is given by
\begin{eqnarray}	
\Gamma_{col}=\frac{p}{k_BT}\sigma_k \bar v
\end{eqnarray}
where $p$ is the gas pressure, $T$ is the temperature and $\bar v$ is the average velocity of atoms and $\sigma_k$ is the kinetic cross section. At room temperatures, the collisional broadening for Rb-Kr is $\Gamma_{col}^{Rb-Kr}=17.1$ MHz/Torr and for Rb-Ne is $\Gamma_{col}^{Rb-Ne}=9.84$ MHz/Torr \cite{buffer-col:JQSRT}. We use 0.5-1 Torr of Krypton as buffer gas in our experiment due to its similar mass to Rb. 
 
   \subsection{Atomic diffusion in a gas cell}

At low buffer gas pressures, a major relaxation process appears due to spatial diffusion out of the laser beams. In our case, we use a cylindrical geometry with transverse diameter of a few millimeters. We are interested in how long it takes for an atom to diffuse out of the transverse cross-section (i.e.  the detection mode volume). 

In two dimensions, the standard deviation in position $\sigma_r$ (defined such that approximately 32\% of atoms
have moved a distance greater than $\sigma_r$) is given by \cite{Vanier:IOP:diff}
\begin{eqnarray}
\sigma_r=2\sqrt{2Dt}
\end{eqnarray}
where $D$ is a diffusion coefficient with units of cm$^2$/s that is defined as
\begin{eqnarray}
D=D_0(\frac{P_0}{P})
\end{eqnarray}
Here $P_0$ = 760 Torr, $P$ is the buffer-gas pressure in Torr, and $t$ is the time. The diffusion constant $D_0$ for Rb-Kr is calculated at temperature of 45$^{\circ}$ to be about 0.16 cm$^2/$sec \cite{Happer:const:1972}. The diffusion constant above room temperature is scaled with $T^{3/2}$.

Longitudinal diffusion could also play a role, especially considering the longitudinal frequency gradient. If an atom moves longitudinally by a distance $\delta z$ during the storage time, it will experience a frequency shift of $\eta\delta z$ at read out. This means that longitudinal diffusion will lead to random frequency shifts of the atoms if the applied gradient is sufficiently large. The effect of longitudinal motion in our experiment is negligible because $\eta\delta z$ is small compared to the unbroadened Raman line.

    \section{Experimental setup and results}
    
\begin{figure}[t]
  \centering\includegraphics[width=\columnwidth]{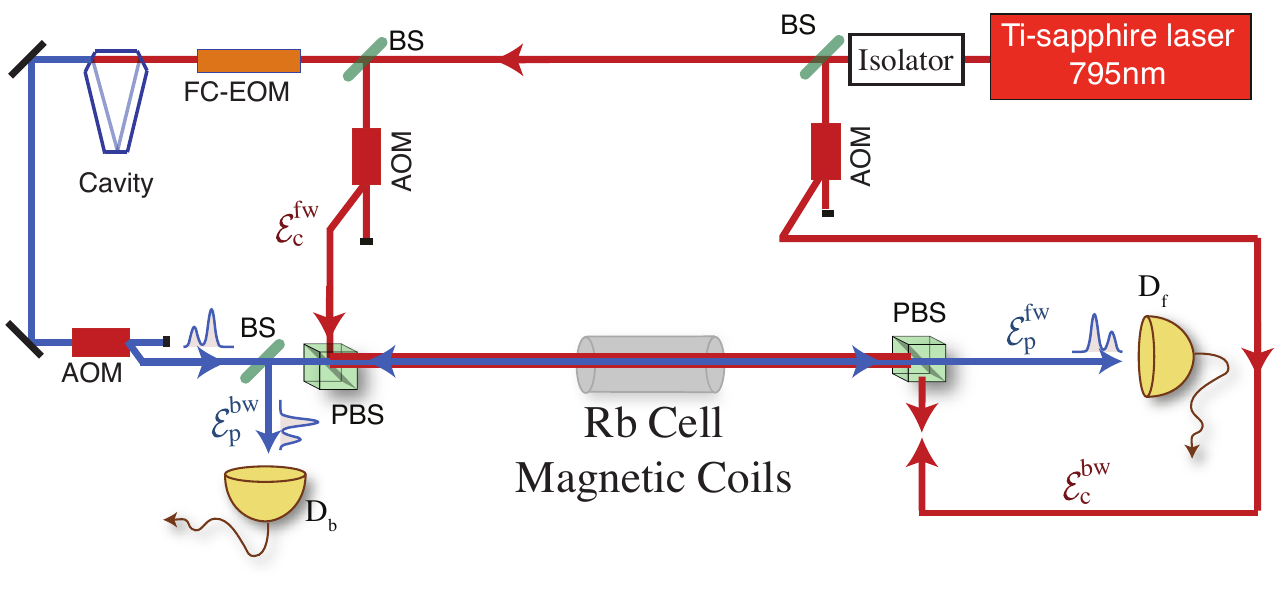}
 \caption[Experimental setup for producing backward propagating echo]{Schematic experimental setup for forward and backward retrieval of echo signal from the memory. The probe field is shifted by 6.8 GHz using a fiber coupled electro-optic modulator (FC-EOM) to match the frequency splitting between the hyperfine ground states of $^{87}$Rb. A cavity was then used to separate the carrier and -6.8 GHz sideband from the probe field after the FC-EOM. BS: beam splitter, PBS: polarized beamsplitter, AOM: acousto-optic modulator, $\mathcal{E}_p^{bw}$: backward-propagating probe field, $\mathcal{E}_p^{fw}$: forward-propagating probe field, $\mathcal{E}_c^{bw}$: backward-propagating coupling field, $\mathcal{E}_c^{fw}$: forward-propagating coupling field, $D_{b/f}$: detectors for the forward and backward-propagating modes.}
  \label{BWsetup}
 \end{figure}

    The results presented in this section were obtained using a $^{87}$Rb mixed with 1 Torr of Kr buffer gas in a warm cell with length of 7.5 cm and diameter of 2.5 cm. The experiment setup is shown in Fig.~\ref{BWsetup}. The probe and coupling fields were collimated with 3 mm and 5 mm diameter, respectively, and had orthogonal linear polarization. A second counter-propagating coupling field was used to retrieve the echo in the backward direction. The Raman detuning was around 1 GHz and the power of the probe and the coupling fields were typically around 10 $\micro$W and 200 mW, respectively. To generate the linear field gradient we used two specially designed solenoids and, by switching the current between them, the sign of the gradient is switched \cite{Hosseini:NComm:2011}. We used two layers of $\mu$-metal shielding to reduce the background magnetic field.    
    
The coupling field can drive the population out of the level $|1\rangle$ or $|2\rangle$ and cause spontaneous Raman scattering (SRS). This can reduce the coherence time of the memory. The loss due to the scattering is given by $\Gamma_{scatt}=\gamma \sqrt{1+\Omega^2 /(\gamma^2+\Delta^2)}$. At large detunings, the scattering rate follows the laser power. At low coupling field powers the FWHM of Raman line is the ground state linewidth $\gamma_0$. We observed that the switching off the coupling field during the storage time increases the coherence time by almost an order of magnitude \cite{Hosseini:NComm:2011}.

        \subsection{Results}
        
   \subsubsection{Shape mirroring}

\begin{figure}[!t]
  \centering\includegraphics[width=\columnwidth]{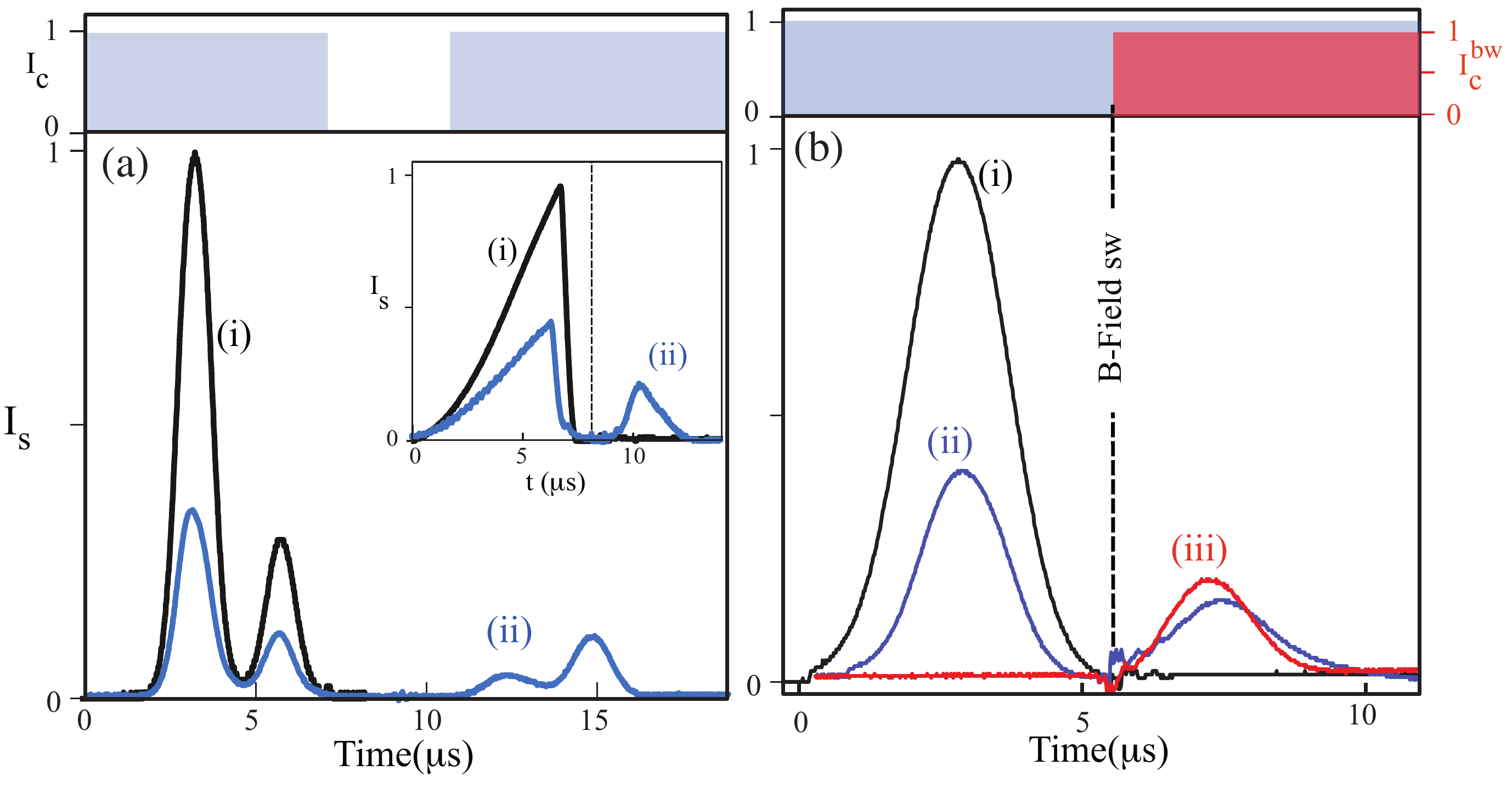}
 \caption{ (a) The shape mirrored photon-echo for a train of two pulses with different amplitude ratio. Inset shows a ramp-shape input (i) and shape mirrored photon echo (ii). Top section shows the switching protocl for the coupling fields. (b) Forward and backward retrieval of echo. (i): The input pulse. (ii): Forward propagating signal light at the output of the memory when the coupling and signal beams are co-propagating. (iii): Backward propagating light at the output of the memory when a coupling field, counter-propagating with respect to input light, is applied at the retrieval stage. Backward propagating light measured using a 50/50 beam splitter before the memory, for this reason trace (iii) is amplified by a factor of 2. Top section shows the switching protocol for the forward (blue) and backward (red) propagating coupling fields. Due to the low OD, part of the input light leaks through the memory without absorption that can be seen in both (a) and (b).}
  \label{bwexp}
 \end{figure}

 One of the intrinsic properties of GEM is that the recalled photon echo has the same shape as the input signal but time reversed. Fig.\ref{bwexp} (a) shows the experimental proof of the shape mirroring for a double Gaussian and ramp shape input pulse, respectively. Output echoes are indeed a mirror image copy of the input pulses due to the time-reversed nature of the process. 

\subsubsection{Backward retrieval of echo}
  
  To experimentally demonstrate backward retrieval of the echo, we used a coupling field propagating in opposite direction with respect to the input pulse. Initially, the input pulse is stored with a co-propagating coupling field. At the read out stage the photon echo was retrieved either in forward or backward direction by means of forward or backward coupling field after the gradient flip. The results are shown in Fig.~\ref{bwexp}(b) where photon echoes retrieved in the forward (trace (ii)) or backward (trace (iii)) directions.
 
    \section{FWM and Raman absorption in a dense atomic sample}
 
The FWM phenomena in EIT systems was studied in Ref. \cite{Lukin:PRL:1997} and experimentally observed later in a dense atomic sample \cite{Phillips:2008p9536, Phillips:2009p11917}. It was shown that FWM in EIT can limit the storage efficiency at high optical depth \cite{Phillips:2008p9536}. In this case, the conversion of an original pulse from the signal to Stokes channel may reduce the read-out efficiency \cite{Matsko:OptExp:2005}. Under certain conditions, however, FWM may lead to gain for both the signal and Stokes fields, which could compensate for any optical losses \cite{Camacho:NPhot:2009}, albeit with added noise.

\begin{figure}[!b]
  \centerline{\includegraphics[width=0.4\columnwidth]{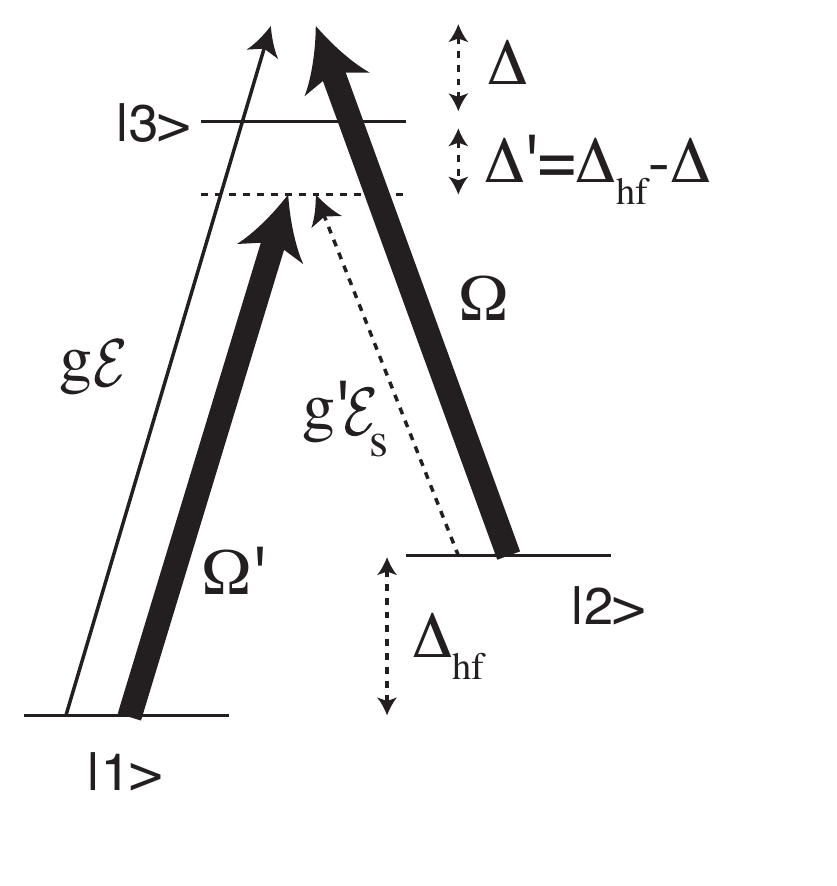}}
    \caption{Atomic level diagram for degenerate FWM between coupling and probe fields. The Rabi frequency of the coupling field for different transitions are shown by $\Omega$ and $\Omega'$ due to different dipole moments. The Rabi frequency of the probe and Stocks fields are shown by $g\mathcal{E}$ and $g'\mathcal{E}_s$, respectively.}
  \label{FWMLS}
  \end{figure} 
  
 With high OD, FWM phenomena can also be observed in a GEM. We consider the level scheme shown in Fig.~\ref{FWMLS} to derive the equations of motion. Assuming a linearly varying two-photon detuning $\delta(z)=\eta z$ and a far-detuned Raman transition, $\Delta\gg\gamma$ we can arrive at the following equation for probe ($\mathcal{E}$) and Stokes ($\mathcal{E}_s$) fields at the centre frequency
\begin{eqnarray}
\frac{\partial}{\partial z}\left(\begin{array}{c} \mathcal{E}(z) \\  \mathcal{E}^*_s(z)\end{array}\right)=i a_0\left(\begin{array}{cc}a_{11} & a_{12} \\a_{21} & a_{22}\end{array}\right)\left(\begin{array}{c} \mathcal{E}(z) \\  \mathcal{E}^*_s(z)\end{array}\right)
\end{eqnarray}
\begin{eqnarray}
a_0&=& \frac{N}{\gamma c(\Omega^2+\Gamma\Gamma_0)}\nonumber \\
a_{11}&=&i g\Gamma_0\nonumber \\
a_{12}&=&a_{21}=-\frac{g'\Omega \Omega'}{\Delta'}\nonumber \\
a_{22}&=&-i\Gamma\frac{g'\Omega'^2}{\Delta'^2}
\label{eq:as}
\end{eqnarray}
and
\begin{eqnarray}
\Gamma_0&=&(\gamma_0 + i \delta) \nonumber \\
\Gamma&=&(\gamma +i(\Delta + \delta ))
\end{eqnarray}
We solve these equations numerically in different regimes to understand the contribution of the FWM process in a Raman absorptive medium. 

\begin{figure}[!h]
  \centerline{\includegraphics[width=\columnwidth]{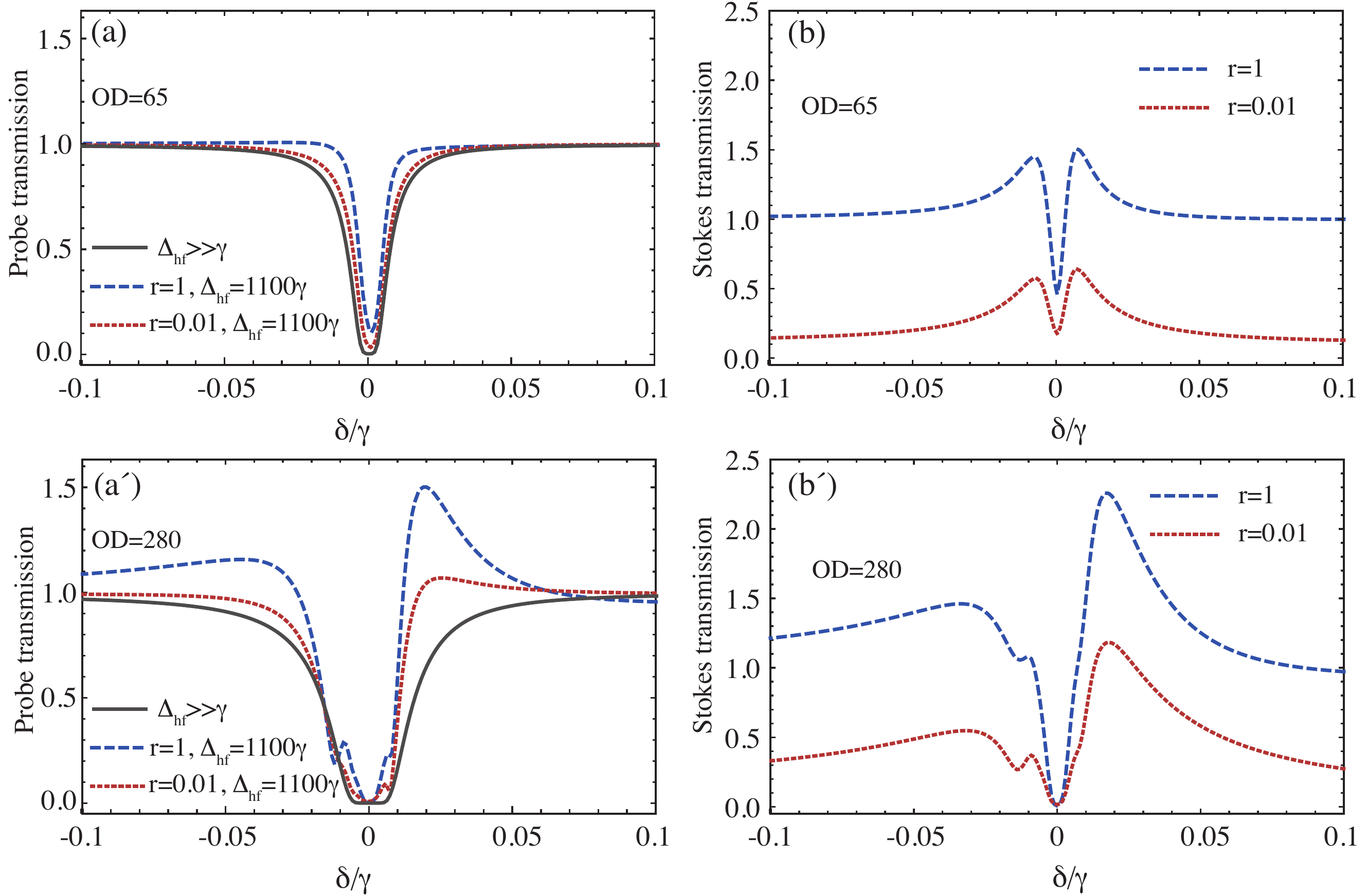}}
    \caption[FWM in GEM medium]{ Transmitted amplitude of, (a)-(a$'$) probe, (b)-(b$'$) Stokes fields normalized to the input probe amplitude for two different ODs. A transmission greater than 1 indicates gain in the medium.  Parameters used for this numerical simulations are: $\Gamma_{p}=4\gamma$, $\gamma_0=0.002\gamma$, $\Omega=3\gamma$, $OD=g^2NL/\gamma c=65,280$, $r=\mathcal{E}_s(z=0)/\mathcal{E}(z=0)=1.0, 0.01$ and $ \Delta=200\gamma$. In the limit $\Delta_{\rm hf}\gg \gamma$ we can effectively switch off the FWM process. In these calculations, the Doppler effect has not been included in our model}
  \label{4WMGEM}
  \end{figure} 
  
Figure \ref{4WMGEM} shows the calculated transmission of the probe and Stokes fields for two different values of resonant OD$=g^2NL/\gamma c$ and two different values of the initial Stokes field amplitude, parameterized by $r=\mathcal{E}_s(z=0)/\mathcal{E}(z=0)$. In our experiment, a non-zero Stokes seed is a possibility.  The probe field is formed by one sideband of a phase modulation that is filtered by one or more resonant cavities.  If the leakage of these cavities is too high, the other sideband of the phase modulation can form a Stokes seed. It can be seen that for large OD, probe  amplification occurs around the edges of the Raman line. The results show that 4WM amplification of the probe can be suppressed by minimizing the seed light of the Stokes field (making $r$ small) and by using an atomic system where $\Delta_{\mathrm{hf}}\gg \gamma$.	
The amplitude of the Stokes field, as well as the gain, is also substantially reduced at the point of maximum probe absorption. The Stokes field is also absorbed around two-photon resonance. FWM inside a vapor cell at 140$^\circ$C has been experimentally investigated in a detuned double-$\Lambda$ configuration  (with $\Delta=150\gamma$) \cite{Boyer:PRL:2007}, where the maximum gain was observed 20 MHz away from the Raman resonance. In a dense atomic medium and at Raman resonance, the competition between large amplification and large absorption leads to complex dynamics which can result in a breakup of the probe pulse. The interference between the probe and Stokes transition can be seen in Fig.~\ref{4WMGEM} (a$'$). 

\begin{figure}[]
  \centerline{\includegraphics[width=0.6\columnwidth]{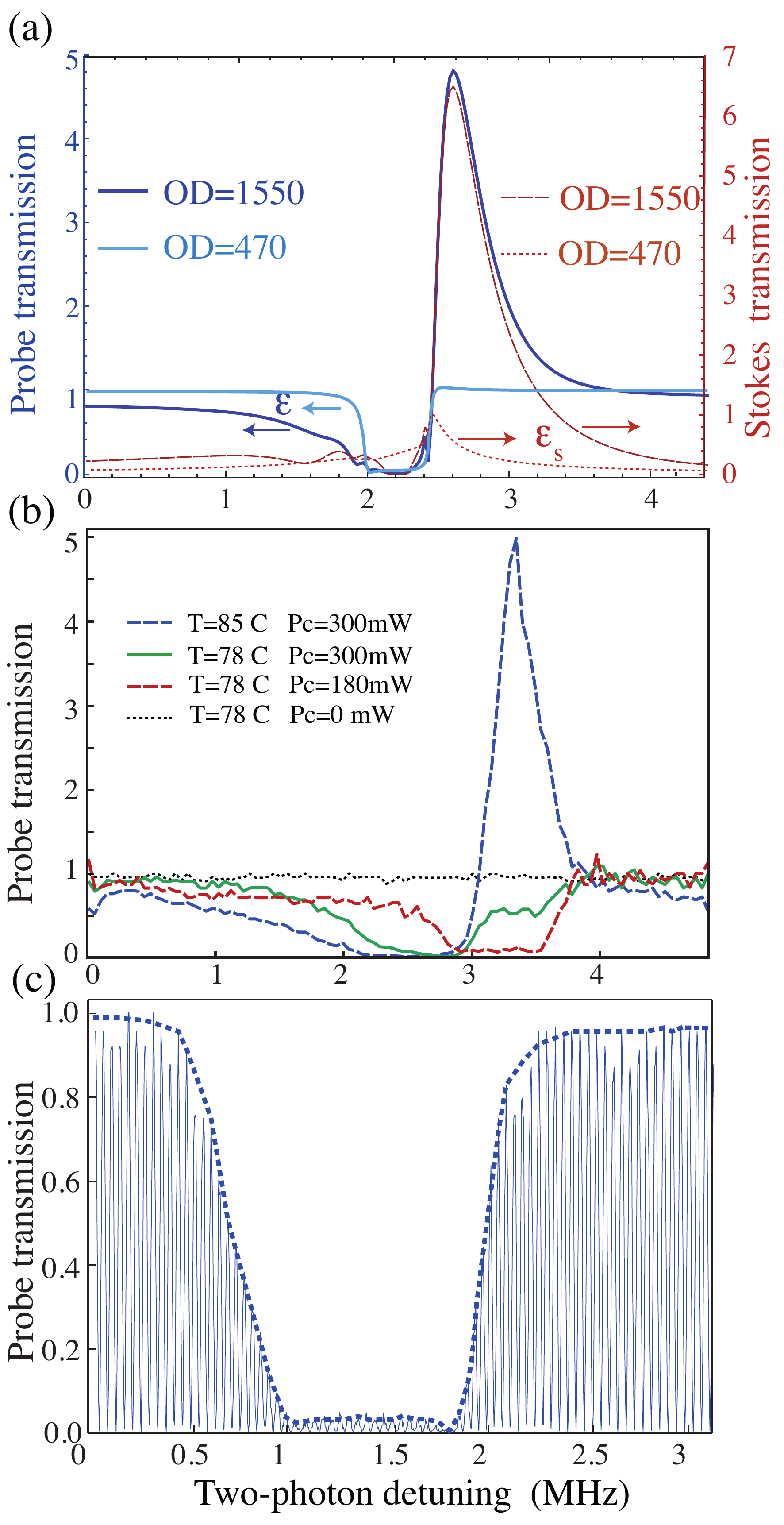}}
    \caption[polarization effect on FWM]{ (a) Transmitted amplitude of the probe and Stokes fields normalized to the input probe amplitude. Parameters used for these numerical simulations are: $\gamma_0=0.002\gamma$, $\Omega=3$ and $5.2 \gamma$, $g^2NL/\gamma c=470,1550$, $r=0.01$, $ \Delta=200\gamma$ and $\eta L=0.08\gamma$, $T=85^o$ C. In these calculations, the Doppler effect has not been included in the model. (b) Experimental measurements of the probe transmission for different temperatures and coupling field power obtained using linear polarization. Absorption is maximum at a non-zero two-photon detuning because of the DC offset magnetic field. (c ) The heterodyne signal showing the broadened Raman absorption line for the same circular polarization of the pump and probe fields with pump power of 350mW.  The probe power in all cases was $\sim$ 1 $\micro$W. The envelope shown is a guide to the eye.}
  \label{4WMGEM-Pol}
  \end{figure}

Using linear polarization instead of circular polarization can significantly enhance the FWM process in a Rb vapor cell. Linear polarization is treated by the atoms as a superposition of left and right circular polarization and will therefore interact with multiple excited state sublevels. This will effectively increase the atom-light coupling strength that enhances the non-linear process of FWM. Fig.~\ref{4WMGEM-Pol} (a) shows results of numerical calculations for transmitted probe and Stokes signals for two different ODs assuming co-propagating beams.

Experimental results showing the broadened Raman lines for different coupling field powers and temperatures are shown in Fig.~\ref{4WMGEM-Pol} (b). These experimental results have been obtained using a 20 cm gas cell containing $^{87}$Rb and 0.5 Torr of Kr buffer gas, while the coupling and probe signal fields have orthogonal linear polarization.  The probe measurement was performed using heterodyne detection. As shown in Fig.~\ref{BWsetup} the probe field is one of the 6.8GHz phase modulation sidebands generated using a fiber coupled electro-optic modulator (FC-EOM). A cavity of finesse $\sim$ 100 was used to select out the probe frequency. Nevertheless, a small amount of seed light (from the other phase modulation sideband) will leak through and enhance the amplification process. Using circular polarization, we can achieve strong absorption of the probe with no measurable gain, as shown in Fig. \ref{4WMGEM-Pol} (c). In this case we have also used two optical cavities for the probe beam, each with a finesse of about 100 providing suppression of the seed signal of more than $10^{6}$ times.

Applying the detuning gradient increases the absorption bandwidth for the probe and therefore widens the region around two-photon resonance in which gain is substantially suppressed and the memory can operate noiselessly \cite{Hosseini:Nphys:2011}. At regimes of large OD applying a gradient does not noticeably reduce the maximum absorption of the probe and it causes suppression of the gain and generated Stokes field. We note here that, in the experiment configuration, all of the beams are co-propagating together and therefore the phase matching condition is satisfied for FWM process. However, the Raman gain can be suppressed if the two $\Lambda$ transitions destructively interfere. The phase of the seeded Stokes field in the experiment is not controlled and we believe that the observed gain is largely due to the vacuum seed. In our model, however, we consider phase of the seeded Stokes field such that the gain is maximum. 

\begin{figure}[!h]
  \centerline{\includegraphics[width=0.6\columnwidth]{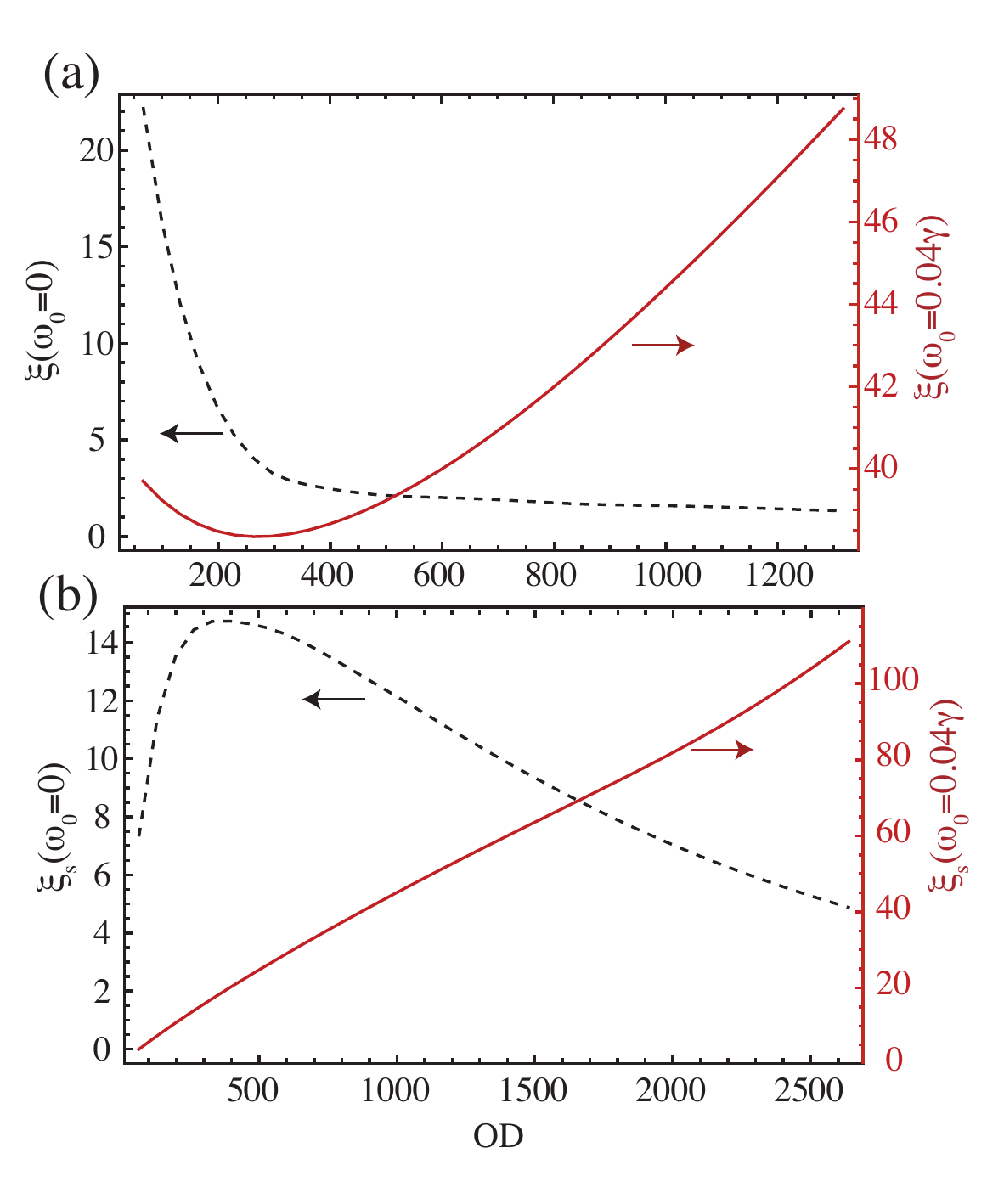}}
    \caption[Probe and Stokes field inside broadened Raman line]{Total (a) probe and (b) Stokes energy within a frequency window of $2\delta=0.04\gamma$ around zero two-photon detuning (dashed black line) and $\omega_0=0.04\gamma$ away from the two-photon resonance (solid red line) calculated for different OD. This quantity corresponds to the total probe and Stokes field energies at the output of the medium. Both axes are normalized to the input probe energy. Parameters used for this numerical simulations are: $\gamma_0=0.002\gamma$,  $ \Delta=200\gamma$, $\Omega=3\gamma$, $r=0.01$, $\eta L=0.08\gamma$. In these calculations, the Doppler effect has not been included in the model.}
  \label{EsvsOD}
  \end{figure}

We calculate the total energy of the Stokes field around two-photon resonance and also $0.04\gamma$ from resonance (where the amplification is maximum) for different optical depths in an inhomogenously broadened medium. As it can be seen in Fig.~\ref{EsvsOD}(a) and (b), the energy of the probe and of the Stokes field around the two-photon detuning,  $\xi(\omega_0=0)=\int_{-\delta}^{\delta}{\mathcal{E}(\omega-\omega_0)d\omega}$, resonance rapidly decreases by increasing the optical depth of the medium. However, $0.04\gamma$ away from the resonance the probe and Stokes energy,  $\xi(\omega_0=0.04\gamma)=\int_{-\delta}^{\delta}{\mathcal{E}(\omega-\omega_0)d\omega}$, increases with OD.  

The significant reduction of Stokes field around the centre of the Raman line indicates large suppression of the FWM process and therefore lack of amplification in this region. The situation is reversed in a dense EIT medium, where the probe transmission is enhanced.
 
\section{Summary}
In summary, we have discussed properties of $\Lambda$-GEM in a warm vapor cell and presented experimental results of shape mirrored and backward retrieved echo. We proposed a scheme based on counter-propagating coupling fields to observe stationary light effect in GEM system. We also discussed the effect of FWM in GEM storage process and interplay between these two processes in a dense atomic medium.

  This research was conducted by the Australian Research Council Centre of Excellence for Quantum Computation and Communication Technology (project number CE110001027).
  
  \section*{References}
  

 \end{document}